# Sub-femtosecond electron bunches created by direct laser acceleration in a laser wakefield accelerator with ionization injection

N. Lemos[1,a)], J. L. Shaw[1], K. A. Marsh[1] and C. Joshi[1]

[1]*University of California Los Angeles, Los Angeles CA USA 90095*

[a)]*Corresponding author: nuno.lemos@ucla.edu*

**Abstract.** In this work, we will show through three-dimensional particle–in-cell simulations that direct laser acceleration in laser a wakefield accelerator can generate sub-femtosecond electron bunches. Two simulations were done with two laser pulse durations, such that the shortest laser pulse occupies only a fraction of the first bubble, whereas the longer pulse fills the entire first bubble. In the latter case, as the trapped electrons moved forward and interacted with the high intensity region of the laser pulse, micro-bunching occurred naturally, producing 0.5 fs electron bunches. This is not observed in the short pulse simulation.

## INTRODUCTION

In the nonlinear laser wakefield accelerator (LWFA) regime [1], as an intense relativistic (> $10^{18}$ W/cm$^2$) short (<50 fs) laser pulse is focused into an underdense plasma (typically $10^{19}$ cm$^{-3}$ for an 800 nm laser beam) the ponderomotive force of the laser radially blows out the plasma electrons, leaving an ion cavity behind. The Coulomb force of the stationary plasma ions then pulls back the blown out electrons back towards the laser axis behind the laser pulse. The resulting wakefield has a periodicity of ~$\lambda_p$ (plasma wavelength) and a phase velocity close to the speed of light, ideal for interacting with particles over a long distance. Moreover, the longitudinal electric field of the wake can have accelerating gradients in excess of 100 GV/m, three orders of magnitude larger then the conventional RF accelerators (100MV/m). Once the bubble is formed, it is necessary to inject electrons into the accelerating electric field. Electrons need to have sufficient initial energy so that they can be trapped and accelerated in the plasma wave. There are several possible injection mechanisms [2-8], but in the ideal nonlinear blowout regime, the laser pulse drives a large enough amplitude wake that some of the plasma electrons blown out by the laser pulse can be injected into the rear of the first wake oscillation through transverse breaking of the plasma wave [2]; this scheme is known as self-injection. These electrons can then "surf" the wake, gaining energy from the longitudinal accelerating field in the bubble. In addition, since the typical scale length of the wakefield is $\lambda_p$ (10 μm for plasma densities of $10^{19}$ cm$^{-3}$), the accelerated electrons eventually bunch with a length shorter then $\lambda_p/4$ (3 fs), which is approximately the length of the accelerating phase of the longitudinal electrical field.

Experiments have shown that LWFAs can produce, in a few millimeters, electron bunches with energies over a GeV [9-12] with, tens of picocoulombs of charge, kiloamps peak current and an inferred bunch duration of a few femtoseconds. Both simulations and theory suggest that such electron beams produced by LWFAs may be good candidates for driving compact free electron lasers (FEL). Experimental effort is ongoing to characterize electron beams from LWFAs, especially their temporal structure, which is an essential parameter for FELs [13]. Experimental studies show that LWFAs can produce electron beam bunches as short as 1.4 fs [14] in the first plasma period. Additionally it was observed that there is also some charge trapped in the subsequent plasma periods forming a train of electron bunches [15], where each bunch is separated by $\lambda_p$ with similar bunch durations. It was also shown that when electrons that are being accelerated in the first plasma period start to dephase (overrun the accelerating field), they can overlap the blue-shifted back part of the laser pulse and start to have density and momentum modulations that lead to micro-bunching [16]. Even though this is a way of producing even shorter electron bunches, micro-bunching in this case only occurs when the electrons already lost most of their energy back to the wake.

In this work, we will show through three-dimensional (3D) particle–in-cell (PIC) simulations that direct laser acceleration (DLA) in conjunction with LWFAs can, lead to microbunching of the trapped electrons while they are in the accelerating phase of the wake, to generate sub-femtosecond electron bunches. In addition, using ionization trapping, this process can be optimized by placing charge in the correct phase of the plasma accelerator structure such that it immediately starts bunching a immediately after trapping.

## IONIZATION INJECTED LWFA IN THE PRESENCE OF DLA

Recent simulations of LWFAs [17] have shown that in addition to wakefield acceleration, LWFAs can transfer energy from the laser to the electrons through a process known as DLA [18-19], which occurs when there is a considerable overlap between the trapped accelerated electrons and the laser field. Not only can this physical process enhance the final energy of the accelerated electrons and their betatron radiation [20], but it can also generate micro-bunched attosecond electron beams before dephasing occurs.

In an ideal LWFA (with fully matched conditions) [1], the laser pulse only occupies the front half of the bubble and the trapped electrons will start to be accelerated at the back of the bubble. These trapped electrons then start to undergo betatron oscillations in response to the linear transverse focusing force in the ion column solely gaining energy from the moving longitudinal field until dephasing. In a non-ideal LWFA where the pulse duration is not matched to the plasma period, it is possible for the laser pulse to occupy the entire bubble. In this case, the trapped electrons that undergo betatron oscillations in the polarization plane of the laser will see an additional electric field from the laser. This transverse electric field of the laser, when in resonance with the betatron motion of the electrons [18-19], will in turn increase the transverse momentum of the trapped electrons. The resonance condition is achieved when a harmonic, N, of the betatron frequency, $\omega_\beta$, is on average [21] equal to the downshifted laser frequency $(1-v_\parallel/v_\Phi)\omega_0$ witnessed by the trapped electrons, where $v_\parallel$ is the longitudinal velocity of the electrons, $v_\Phi$ is the phase velocity of the laser and $\omega_0$ is the laser frequency. The increase in transverse momentum can then be converted into longitudinal momentum via the v x B force, and this process is known as DLA. Additionally, ionization injection can be used to advantageously produce the charge within the wake such that it overlaps the laser pulse at the moment of trapping.

## 3D OSIRIS SIMULATIONS OF A LWFA WITH DLA

In order to study electron bunching in LWFAs, two 3D simulations were done using the PIC code OSIRIS 2.0 [20] with the two different laser pulse durations, such that the shortest laser pulse (25 fs) occupies only a fraction of the first bubble, whereas the longer laser pulse (45 fs) fills the entire first bubble. The simulations have been carried out in the speed of light frame within a box that was 66 x 98 x 98 µm with 2550 x 652 x 652 number of cells. The longitudinal resolution was $\lambda_0/30$ and both transverse directions had a resolution of $\lambda_0/5.4$, where $\lambda_0$ is the laser wavelength. The layout of the 3D simulations is shown in Figure 1. The moving window containing the laser pulse is launched from the left hand side into a region filled with neutral gas that comprised 99% helium and 0.1% nitrogen. The gas profile had a 100 µm density up ramp followed by a 1000 µm constant density region and a 100 µm density down ramp.

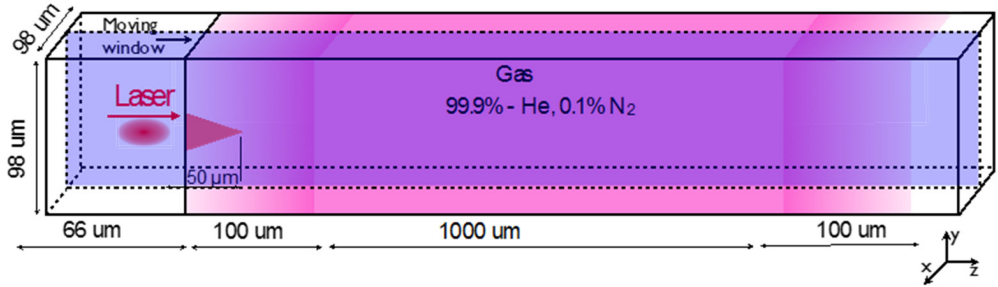

**Figure 1 –** Layout of the 3D PIC simulation in the speed-of-light frame within a box that was 66 x 98 x 98 μm with 2550 x 652 x 652 number of cells. The laser had an $a_0$ = 2.1 and was focused 50 μm into the 100 μm density up ramp in a region of gas that comprised a mix of 99.9% He and 0.1% of $N_2$. The density ramp was followed by a 1000 μm constant density region ending in a 100 μm down density ramp

The ionization of the K-shell electrons of the nitrogen served to inject charge into the wake as soon as it is formed (ionization injection), which is predominantly comprised by ionized helium atoms. The laser pulse is linearly polarized (in the yz plane) with $\lambda_0$ = 815 nm and it is focused in the middle of the left density ramp to a spot size of 6.7 $\mu m$ (FWHM of the electric field). The pulse duration was 25 fs in the short pulse simulation and 45 fs for the long pulse simulation, while the peak normalized vector potential was $a_0$ = 2.1 for both simulations. The short pulse simulation acted as the control case simulation where the laser pulse did not interact with the trapped electrons and no DLA and bunching was expected. The long pulse simulation models the case where the back of the laser pulse overlaps the trapped electrons in order for DLA and bunching to occur.

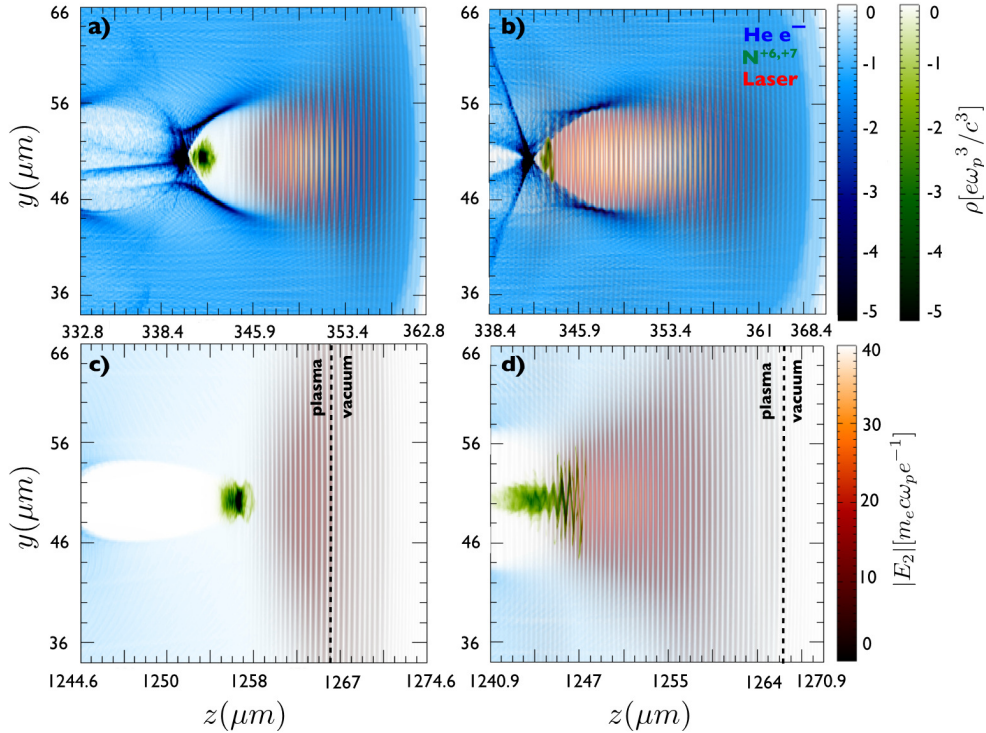

**Figure 2:** Snapshots of the electron charge density of the central plane in the 3D simulations of $He^{+1,+2}$ and $N^{+1,+5}$ (in blue) and $N^{+6,+7}$ (in green) for the two simulations. The top row is 1ps into the simulation and the bottom row is 4 ps into the simulation. The left column presents the simulation with a laser pulse of 25 fs, and the right column presents results for a pulse of 45 fs. The absolute value of the laser electric field is represented in red. The dashed black line indicates the plasma vacuum boundary..

Figure 2 shows the results of two full-scale 3D simulations of a LWFA. The central slice (shown as a blue plane in figure 1) of the wakefield structure is shown for the 25 fs pulse duration simulation at 1ps (a) and 4 ps (c) into the simulation and for the 45 fs pulse duration for the same times (b) and (d) respectively. The blue color bar represents the electron charge density for the plasma comprised of $He^{+1,+2}$ and $N^{+1,+5}$, the green color bar represents the charge density of the trapped electrons from $N^{+6,+7}$ and the red color bar represents the absolute value of laser electric field. As the laser propagates, the gas is readily ionized via tunnel ionization and the ponderomotive force of the laser expels the $He^{+1,+2}$ and $N^{+1,+5}$ electrons from the laser axis forming the blowout region and the accelerating structure (white region in fig.2 (a) and (b)). The K-shell electrons from nitrogen ($N^{+6,+7}$) are ionized close to the peak of the laser beam and injected into the accelerating structure. Those electrons that gain enough energy - so as to move with a velocity equal to the phase velocity of the wake in the lab frame - from the longitudinal electric field as they slip back in the wake are trapped in the first buckle (green electrons in fig.2). Notice the absence of any blue charge in the wake meaning that there is no self-trapping of electrons in either simulation. In the 25fs pulse duration simulation the trapped electrons continuously gain energy from the longitudinal electric field until the end of the simulation (fig2.(c)), never overlapping with the back part of the laser pulse, as is expected for a matched LWFA. In the case of the 45 fs pulse duration simulation, as soon as the electrons are trapped they overlap the back of laser pulse (fig.2(b)). The electrons that undergo betatron motion in the plane of polarization of the laser pulse and meet the quai-resonance condition will also gain energy from DLA [21]. As the electrons move forward and interact with the high intensity region of the laser pulse, micro-bunching occurs naturally as in an inverse free electron laser (IFEL). Figure 2 (d) shows the trapped electron bunch at the end of the simulation, where it is possible to see that the electrons that interacted with the laser pulse are micro-bunched and the ones that did not interacted with the laser pulse maintain their original shape. The period of the betatron oscillations is not uniform in space (as in an IFEL) and when in resonance, it only dependents on the laser. With a linear polarized laser beam, the transverse velocity of the resonant electrons oscillates with the laser period, while the longitudinal velocity oscillates twice per laser period [18]. This in turn leads to a natural bunching of the trapped electrons at half the laser wavelength.

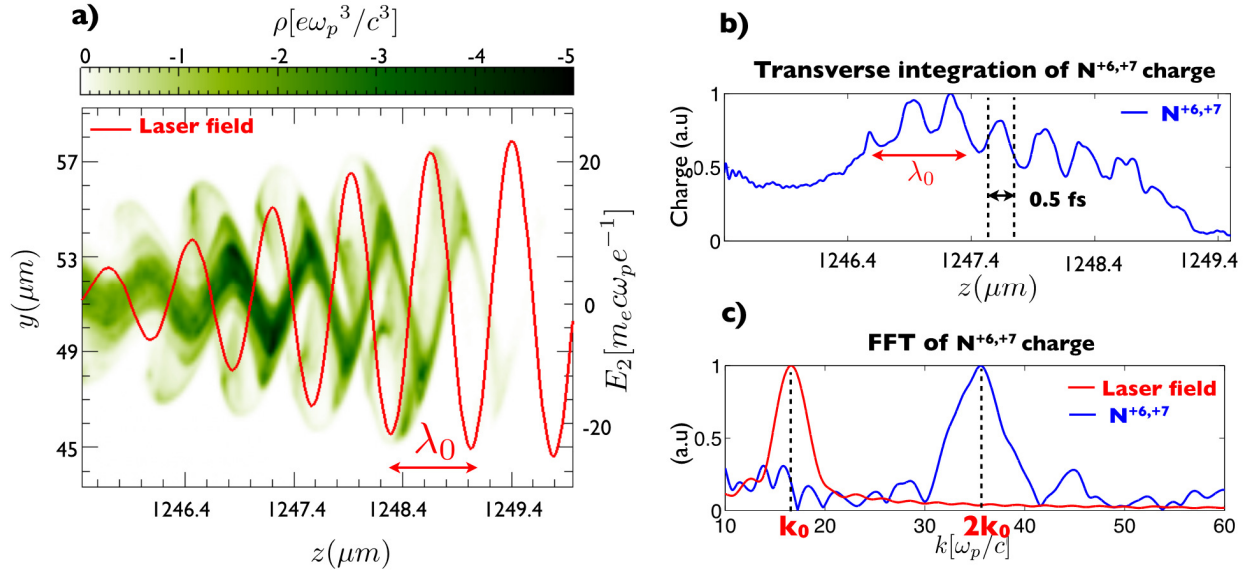

**Figure 3:** (a) $N^{+6,+7}$ trapped electrons charge density and the lineout of the laser electric field in the center of the simulation. (b) Integration in the y axis of the electron charge density in (a). (c) Fourier transform of (b) in blue and of the laser field in red.

Figure 3 (a) shows a magnification of fig. 2(d) where the trapped electrons overlap the back of the laser pulse (red line) and naturally bunch at twice the laser wavelength. Integrating the density charge in fig. 3(a) in the y direction (see fig. 3(b)), it is possible to identify the electron bunching structure and each bunch has an average duration of 0.5 fs. The Fourier transform of fig. 3(b) is shown in fig. 3(c) in blue where the main frequency is $2k_0$, where $k_0$ is the laser main frequency that overlaps the electrons represented by the red curve. This confirms that there is electrons bunching ate twice the laser frequency.

## SUMMARY

We have shown by 3D PIC simulations with the code OSIRIS 2.0 that in a LWFA with non-matched laser pulse durations, the trapped electrons that overlap the laser field can further increase their energy [17, 22] due to the process of DLA and naturally micro-bunch at half the laser wavelength. Although the ionization-trapped electrons execute betatron oscillations due to the ion column in the long and short pulse simulations, in the long pulse duration simulation, there is significant interaction between the trapped electrons and the laser. Thus leading to DLA for those electrons that undergo betatron motion in the plane of polarization of the laser pulse and obey the resonance condition. These electrons will then micro-bunch with average bunch duration of 0.5 fs. This method can be optimized by using even longer laser pulse durations where all the trapped electrons interact with the laser pulse for a longer time improving the bunching quality.

## ACKNOWLEDGMENTS

Work supported by DOE grant DE-SC0010064. Simulation work done on Blue Waters.